\documentclass[11pt]{article}
\usepackage{moriond,amssymb}
\usepackage{graphicx}

\def\Journal#1#2#3#4{{#1} {\bf #2}, #3 (#4)}

\def\NPB{{\em Nucl. Phys.} B}

\def\PRL{\em Phys. Rev. Lett.}
\def\PRD{{\em Phys. Rev.} D}

\def\JHEP{\em JHEP}

\def\be{\begin{equation}}
\def\ee{\end{equation}}
\def\bea{\begin{eqnarray}}
\def\eea{\end{eqnarray}}

\begin{document}
\vspace*{4cm}
\title{QCD Aspects of the NuTeV Anomaly}

\author{Stefan Kretzer}

\address{Physics Department, Brookhaven National Laboratory,
Upton, New York 11973, USA, and\\
RIKEN-BNL Research Center, Brookhaven National Laboratory,
Upton, New York 11973, USA}

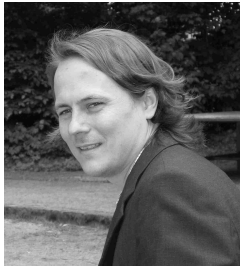
\maketitle

\abstracts{The weak mixing angle $\sin^2 \theta_{\mathrm{W}}$ 
measured in neutrino scattering differs from 
the world average of other 
measurements by about $3 \sigma$. 
I discuss QCD corrections of perturbative and nonperturbative 
(parton structure) origin
to the underlying neutrino observables.}

\section{Introduction}
An important open question in particle physics in recent years has been the
significance of the \textquotedblleft NuTeV anomaly\textquotedblright ---a
$3\,\sigma $ deviation of the measurement of
$\sin ^{2}\theta _{\mathrm{W}}$ ($0.2277\pm 0.0013\pm 0.0009$)
in neutrino scattering~\cite{nutev}, from the
world average of other measurements~\cite{lepew}
($0.2227\pm 0.0004$).
Possible sources of the NuTeV anomaly, both within and beyond the
standard model, have been examined by Davidson {\it et al.}~\cite{forte}.
The NuTeV measurement was based on a
correlated fit to the ratios  $R^{\nu, {\bar \nu}}_{exp}$ of 
``long'' and ``short'' events [dominated by
CC and NC interactions, respectively] 
in sign-selected neutrino and
anti-neutrino scattering on a (primarily) iron target at Fermilab.
Here and below, the subscript {\it exp} will
label observables defined in terms of the experimental
event length, to distinguish them from the (unlabeled)
theory ratios defined in terms of CC and NC interactions.
The differences between the experimentally accessible observables
$R^{\nu, {\bar \nu}}_{exp}$
and the fundamental ratios 
$R^{\nu, {\bar \nu}}$
that can be calculated in theory are significant, 
e.g.~$R^{\nu}$ and $R^{\nu}_{exp}$ differ by $\sim 20\%$ when
the accuracy of  $R^{\nu}_{exp}$ is $<1\%$.
The complexity of the Monte Carlo analysis~\cite{nutev}  
in terms of event length limits the conclusiveness of a theory assessment 
of the anomaly in several respects. 
E.g.~the NuTeV analysis procedure 
of simultaneously fitting $R^{\nu, {\bar \nu}}_{exp}$
is related but not identical 
to measuring the Paschos-Wolfenstein ratio~\cite{pw} 
$R^{-}$ 
\begin{equation}
\label{eq:pw}
R^{-}  \equiv  \frac
{\sigma _{\mathrm{NC}}^{\nu }-\sigma _{\mathrm{NC}}^{\bar{\nu}}}
{\sigma _{\mathrm{CC}}^{\nu }-\sigma _{\mathrm{CC}}^{\bar{\nu}}}
\simeq \frac{1}{2}-\sin ^{2}\theta_{\mathrm{W}}
+\delta R_{A}^{-}\ +\delta R_{QCD}^{-}+\delta R_{EW}^{-}
\end{equation}
where the three correction terms are due to the non-isoscalarity of the
target ($\delta R_{A}^{-}$),
NLO and nonperturbative
QCD effects ($\delta R_{QCD}^{-}$),
and higher-order electroweak effects ($\delta R_{EW}^{-}$).
We can split the QCD corrections
up some more and write
\begin{equation}
\delta R_{QCD}^{-}
=\delta R_{s}^{-}+\delta R_{I}^{-}+\delta R_{NLO}^{-}
\label{eq:QCD}
\end{equation}
where the three terms on the right-hand side are due to possible strangeness
asymmetry ($s^-=s-\bar{s}\neq 0$) and isospin violation ($u_{p,n}\neq
d_{n,p}$) effects in the parton structure of the nucleon,
and NLO corrections, respectively. 
The
original NuTeV analysis was carried out at LO in QCD and assumed $\delta
R_{s}^{-}=0=\delta R_{I}^{-}$.

\section{Perturbative Corrections:}
At sufficiently high neutrino energy,
the total neutrino cross section
\begin{equation}
\sigma ^{\nu }\equiv \sigma ^{\nu N\rightarrow lX}
=\int d^{3}p_{l}\ \frac
{d^{3}\sigma ^{\nu N\rightarrow lX}}{d^{3}p_{l}}
\label{eq:sigtot}
\end{equation}
can be calculated in QCD perturbation theory. 
The differential cross section in Eq.~(\ref{eq:sigtot})
factorizes into a sum of convolutions of parton distribution
functions and partonic cross sections
\begin{equation}
d^{3}\sigma ^{\nu N\rightarrow lX}
=\sum_{f=q,g}f\otimes d^{3}\sigma ^{\nu f\rightarrow lX}\ \ .
\label{eq:factorem}
\end{equation}
This calculation has been performed at NLO accuracy~\cite{kr03}.
\begin{table}[t]
\caption{$R^{\nu, {\bar \nu}}$ evaluated in pQCD 
\label{tab:rnu}}
\begin{center}
\begin{tabular}{l|l|l}
PDF ($\sin^2 \Theta_{\rm W}$)  &  $R^\nu$  &  $R^{\bar \nu}$ \\
\hline 
GRV NLO (0.2227) & 0.3120  & 0.3844 \\
GRV LO (0.2227)  & 0.3125 & 0.3860 \\
GRV NLO (0.2277) & 0.3088 & 0.3839 \\
CTEQ6 NLO (0.2227) & $0.3105 
$ 
& 0.$3841 
$
\end{tabular}
\end{center}
\end{table}
The analysis included target and charm mass effects.
Also included are the non-isoscalarity of the target
material (iron), i.e., $\delta R_{A}^{-}$ in (\ref{eq:pw});
energy averaging over the neutrino and anti-neutrino flux spectra;
and cuts in hadronic energy
as used in the experimental analysis~\cite{nutev}.

Table \ref{tab:rnu} summarizes the QCD corrections to $R^{\nu, {\bar \nu}}$. 
(Deviations from moment estimates~\cite{forte,dobel} can
be traced back~\cite{kr03} to the approximations required 
to apply the moment technique.)
The NLO corrections to $R^{\nu,{\bar \nu}}$ are of 
comparable size than the experimental errors assigned to 
$R^{\nu,{\bar \nu}}_{exp}$. 
Because of the above mentioned mismatch between 
$R^{\nu,{\bar \nu}}$ and $R^{\nu,{\bar \nu}}_{exp}$
it is impossible to predict how these corrections
will shift the central value of the
Weinberg angle  $\sin^2 \theta_{\mathrm{W}}$.
The perturbative QCD corrections to $R^{\nu,{\bar \nu}}$
largely cancel in $R^{-}$~\cite{kr03} but it remains unclear
to what extent the NuTeV LO Monte Carlo analysis is equivalent to 
measuring this ratio.

In the context of perturbative corrections it should also be noted that
a reanalysis~\cite{Hollik} of electroweak corrections to $R^{\nu}$
points at potentially significant effects as well.

\section{Nuclear Effects and Higher Twists}
In principle, the parton distribution functions in Eq.~(\ref{eq:factorem})
should be those of nuclear targets. Our calculation is done as an incoherent
sum of contributions from parton densities of unbound nucleons.
Experimental information on nuclear PDFs is relatively scarce, and nuclear
PDFs only account for leading twist 2 ($\tau = 2$) effects. Higher twists,
whether they relate to nuclear modifications or not, are
generally difficult to handle consistently. By limiting ourselves to $\tau =
2$, our error estimates may be underestimates.
Flavour non-diagonal higher twists (``cat's ears'') 
can have an impact on the measurement of the weak mixing angle in neutrino
scattering off a non-isoscalar target \cite{ht}.

\section{Strangeness Asymmetry:}

An asymmetric strange sea in the nucleon ($s^-\neq 0$)
contributes to a correction term to $R^{-}$ at LO.
Neglecting some other effects, we have 
approximately that~\cite{forte}
\begin{equation}
\delta R_{s}^{-} \simeq
- \left(
\frac{1}{2}-\frac{7}{6}\sin^{2}\theta_{\mathrm{W}}
\right)
\frac{[S^{-}]}{[Q^{-}]},
\label{eq:pws}
\end{equation}
where the strangeness asymmetry is quantified by
the second moment integral
\begin{equation}
\left[ S^{-}\right] \equiv
\int x \left[ s(x)-{\bar{s}}(x)\right] dx \; ;
\label{eq:mom2}
\end{equation}
and $[Q^{-}]$
represents the isoscalar up and
down quark combination.
A positive moment $[S^-]$ works in the direction to improve
the anomaly and a value of about half a percent would remove
it, altogehter.
While the first moment
\begin{equation}
\int \left[s(x)-\overline{s}(x)\right] dx = 0 \, ,
\label{eq:strnumsumrule}
\end{equation}
has to be zero as an exact sum rule there is not even an
approximate symmetry 
[CP conjugation turns $s(x)$ into ${\bar s}_{\bar p}(x)$
-- the anti-strange sea of the anti-proton]
that would protect the second moment
$[S^-]$. It is, therefore, bound to be nonzero.
As an even moment of a $q - {\bar q}$
difference $[S^-]$ is not
among the local quark operators that are probed in DIS. 
Being not accessible to the lattice, it affords a phenomenological 
determination. A recent global CTEQ PDF analysis~\cite{DimuonFitting} 
has included data 
(``dimuon events'')~\cite{dimuons}
on the neutrino-
and antineutrino-production of charm 
\begin{eqnarray}
W^+ + s \rightarrow c \\
W^- + {\bar s} \rightarrow {\bar c}
\end{eqnarray}
It finds a central value $[S^-] \simeq 0.002$ and 
conservative bounds
\begin{equation}
-0.001 < [S^-] < 0.004\ \ \ .
\end{equation}
Via the NLO calculation described above, this translates into
\begin{equation}
-0.005 <  \delta (\sin^{2}\theta_{W}) < +0.001\ \ \ .
\end{equation}

The shift in $\sin^{2}\theta_{W}$
corresponding to the central fit bridges a substantial part 
($\sim 1.5 \sigma$) of the original 
$3\,\sigma$ discrepancy between the NuTeV result and the
world average of other measurements of $\sin ^{2}\theta_{\mathrm{W}}$.
For PDF sets with a shift toward the negative end, such as $-0.004$,
the discrepancy is reduced to less than 
$1\,\sigma $.  On the other hand, for PDF sets with a shift toward the
positive end, such as $+0.001$, the discrepancy remains.

\section{Isospin Violations:}

Isospin symmetry violation effects at the parton level contribute 
a shift of $R^-$ that can be gauged by the second moment
$[D^-_N-U^-_N]$, where
$N=(p+n)/2$. 
The MRST collaboration~\cite{mrst} has recently made a first
attempt to separate proton and neutron PDFs where isospin for the valence
quarks is broken by a function with a single parameter $\kappa $.
We have applied the candidate MRST
PDFs~\cite{mrst} to our NLO calculation and find that
the range of allowed $\kappa$ parameter~\cite{mrst}, 
$-0.7<\kappa <0.7$, implies
\begin{equation}
-0.007\lesssim \delta R_{I}^{-}\lesssim 0.007 \ .
\label{eq:mrst}
\end{equation}
The best fit value of $\kappa =-0.2$ corresponds
to a shift of $\delta R_{I}^{-}=-0.0022$, bringing the
anomaly down to $\sim 1.5 \sigma$.

\section{Conclusions:}
The uncertainties in the parton structure of the nucleon that
relate to $R^{-}$ will not decrease substantially any time soon.
The uncertainties in the theory that relates $R^{-}$ to
$\sin ^{2}\theta_{\mathrm{W}}$ are substantial on the scale of
precision of the high statistics NuTeV data.
Within their bounds, the results of this theory
study suggest that
the new dimuon data, the Weinberg angle measurement,
and other global data sets used in QCD parton structure
analysis can all be consistent within the standard model of
particle physics. The central value of
the Weinberg angle measured in neutrino scattering will also depend
on perturbative QCD and electroweak corrections  
to the experimentally observable $\nu$ and ${\bar \nu}$ 
cross section ratios $R^{\nu, {\bar \nu}}_{exp}$
and a definitive statement can only be obtained from an 
experimental re-analysis of the NuTeV data. 
Before a careful re-assessment of all theoretical 
corrections \& uncertainties (pQCD \& non-pQCD \& electroweak), 
the significance of the $\sim 3 \sigma$ discrepancy with the 
standard model remains questionable.

\section*{Acknowledgments}
I thank the organizers of the Moriond 2004 QCD session
for the invitation and Dave Mason for a very cooperative
interaction at the meeting. My participation has been financed by 
RIKEN-BNL and by an NSF subsidy through Brown University. 
The results represent collaborations with P.~Nadolsky, 
F.~Olness, J.~Owens, J.~Pumplin, M.H.~Reno, D.~Stump and W.-K.~Tung; 
supported by DOE contract No.~DE-AC02-98CH10886.

\end{document}